\documentclass[aps,prb,reprint,superscriptaddress,preprintnumber,showpacs]{revtex4-1}
\usepackage{graphicx}
\usepackage{color}
\usepackage{dcolumn}

\begin{document}

%\begin{CJK*}{GB}{}

% Use the \preprint command to place your local institutional report
% number in the upper righthand corner of the title page in preprint mode.
% Multiple \preprint commands are allowed.
% Use the 'preprintnumbers' class option to override journal defaults
% to display numbers if necessary
%\preprint{}

%Title of paper
\title{Quantum transport of two-dimensional Dirac fermions in SrMnBi$_2$}
\author{Kefeng Wang}
\affiliation{Condensed Matter Physics and Materials Science Department, Brookhaven National Laboratory, Upton New York 11973 USA}
\author{D. Graf}
\affiliation{National High Magnetic Field Laboratory, Florida State University, Tallahassee, Florida 32306-4005, USA}
\author{Hechang Lei}
\affiliation{Condensed Matter Physics and Materials Science Department, Brookhaven National Laboratory, Upton New York 11973 USA}
\author{S. W. Tozer}
\affiliation{National High Magnetic Field Laboratory, Florida State University, Tallahassee, Florida 32306-4005, USA}
\author{C. Petrovic}
\affiliation{Condensed Matter Physics and Materials Science Department, Brookhaven National Laboratory, Upton New York 11973 USA}

\date{\today}

\begin{abstract}
We report two-dimensional quantum transport in  SrMnBi$_2$ single crystals. The linear energy dispersion leads to the unusual nonsaturated linear magnetoresistance since all Dirac fermions occupy the lowest Landau level in the quantum limit. The transverse magnetoresistance exhibits a crossover at a critical field $B^*$ from semiclassical weak-field $B^2$ dependence to the high-field linear-field dependence. With increase in the temperature, the critical field $B^*$ increases and the temperature dependence of $B^*$ satisfies quadratic behavior which is attributed to the Landau level splitting of the linear energy dispersion. The effective magnetoresistant mobility $\mu_{MR}\sim 3400$ cm$^2$/Vs is derived. Angular dependent magnetoresistance and quantum oscillations suggest dominant two-dimensional (2D) Fermi surfaces. Our results illustrate the dominant 2D Dirac fermion states in SrMnBi$_2$ and imply that bulk crystals with Bi square nets can be used to study low dimensional electronic transport commonly found in 2D materials like graphene.
\end{abstract}

\maketitle
%\end{CJK*}

Dirac fermions have raised great interest in condensed matter physics, as seen on the example of materials such as graphene \cite{graphene1} and topological insulators (TIs).\cite{ti1} The linear dispersion between momentum and energy of Dirac fermions brings forth some spectacular properties, such as zero effective mass and large transport mobility.\cite{graphene1,ti1} In addition to the surface/interface states in TIs and graphene, Dirac states in bulk materials were discussed in organic conductors~\cite{organic} and iron-based superconductors such as BaFe$_2$As$_2$.\cite{iron1,iron2} Recently, highly anisotropic Dirac states were observed in SrMnBi$_2$,\cite{lattice,srmnbi21} where linear energy dispersion originates from the crossing of two Bi $6p_{x,y}$ bands in the double-sized Bi square nets. \textbf{SrMnBi$_2$ has a crystal structure similar to that of the superconducting Fe pnictides and is a bad metal.\cite{srmnbi21,srmnbi22}} The Fermi velocity along $\Gamma-M$ symmetry line is $\nu _{F}^{\parallel }\approx 1.51\times 10^{6}$ m/s, whereas the Fermi velocity in the orthogonal direction experiences nearly one order of magnitude decrease.\cite{srmnbi21,srmnbi22}

One of the interesting properties of Dirac materials is the quantum transport phenomena.\cite{quantumtransport,quantummr} Unlike  the conventional electron gas with parabolic energy dispersion, where Landau levels (LLs) are equidistant,\cite{metal} the distance between the lowest and $1^{st}$ LLs of Dirac fermions in magnetic field is very large and the quantum limit where all of the carriers occupy only the lowest LL is easily realized under moderate fields.\cite{LL1,LL2} Consequently some quantum transport phenomena such as quantum Hall effect and large linear magnetoresistance (MR) could be observed by conventional experimental methods in Dirac fermion system.\cite{qt1,qt2,qt3,qt4}

Here we show two-dimensional (2D) quantum transport in bulk SrMnBi$_2$ single crystals. The linear energy dispersion leads to the unusual nonsaturated linear MR since all Dirac fermions occupy the lowest LL in the quantum limit. The transverse MR exhibits a crossover at a critical field $B^*$ from semiclassical weak-field MR $\sim$ $B^2$ to the high-field MR $\sim$ B dependence. The critical field $B^*$ increases with the increase in temperature and its temperature dependence satisfies quadratic behavior which is attributed to the Landau level splitting of the linear energy dispersion. Angular dependent MR and oscillation indicates the quasi-2D Fermi surfaces (FS). We derive the effective magnetoresistant mobility $\mu_{MR}\sim 3400$ cm$^2$/Vs. Our results illustrate the dominant 2D Dirac fermion states in SrMnBi$_2$.

Single crystals of SrMnBi$_2$ were grown using a self-flux method.\cite{crystal} Stoichiometric mixtures of Sr (99.99$\%$), Mn (99.9$\%$) and excess Bi (99.99$\%$) with ratio Sr:Mn:Bi=1:1:9 were sealed in a quartz tube, heated to 1050 $^{\circ}C$ and cooled to 450 $^{\circ}C$ where the crystals were decanted. X-ray diffraction (XRD) data were taken with Cu K$_{\alpha}$ ($\lambda=0.15418$ nm) radiation of Rigaku Miniflex powder diffractometer. Transport measurements were conducted in a Quantum Design PPMS-9 with conventional four-wire method. The crystal was cleaved to a rectangular shape with dimension 4$\times$1 mm$^{2}$ in the \textit{ab}-plane and 0.2 mm thickness along the \textit{c}-axis. For in-plane resistivity $\rho_{ab}(T)$, the current path was in the \textit{ab}-plane, whereas magnetic field was perpendicular to the current and parallel to the \textit{c}-axis except in the rotator experiments. The $c$-axis resistivity $\rho_c(T)$ was measured by attaching current and voltage wires to the opposite sides of the plate-like crystals.\cite{caxis1} High field MR oscillation were performed at National High Magnetic Field Laboratory in the same configuration to the in-plane MR.

%%%%%%%%%%%%%%%%%%%%%%% Figure 1 %%%%%%%%%%%%%%%%%%%%%%%%%%%%%%%%%%
\begin{figure}[tbp]
\includegraphics[scale=0.65]{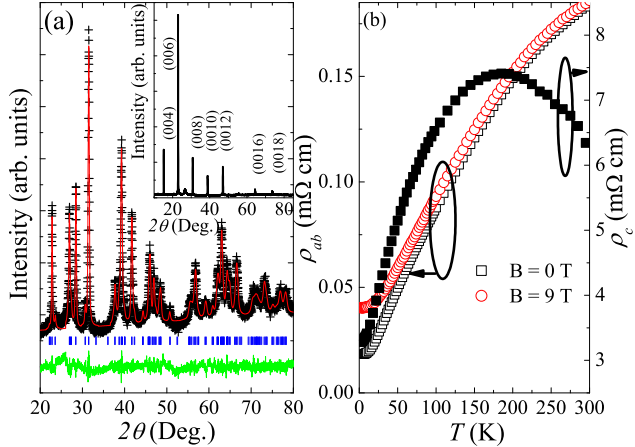}
\caption{(a) Powder XRD patterns and structural refinement results. The data were shown by ($+$) , and the fit is given by the red solid line. The difference curve (the green solid line) is offset. The inset shows the single crystal XRD pattern indicating the c-axis orientation of crystal. (b) Temperature dependence of the in-plane resistivity $\protect\rho_{ab}(T)$ (open symbols) and $c$-axis resistivity $\rho_c(T)$ (filled symbols) in $B=0$ T (squares) and $B=9$ T (circles) magnetic field respectively. }
\end{figure}
%%%%%%%%%%%%%%%%%%%%%%%%%%%%%%%%%%%%%%%%%%%%%%%%%%%%%%%%%%%%%%%%%%%%

All powder and single crystal XRD reflections can be indexed in the I4/mmm space group by RIETICA software (Fig. 1(a)).\cite{rietica} The determined lattice parameters are $a=b=0.4561(8)$ nm and $c=2.309(6)$ nm in agreement with the published data.\cite{lattice} The in-plane resistivity $\rho_{ab}(T)$ shown in Fig. 1(b) exhibits a metallic behavior. An external magnetic field enhances the resistivity. As the temperature is increased, MR is gradually suppressed and is rather small above $\sim 60$ K. Resistivity along the $c$-axis ($\rho_c(T)$) is nearly two orders of magnitude larger than $\rho_{ab}(T)$ and exhibits a weak crossover at high temperature. In what follows we will discuss in-plane MR.

Angular dependent MR $\rho(B,\theta)$ at $T\sim 2$ K is shown in Fig. 2(a)-(b). The crystal was mounted on a rotating stage such that the tilt angle $\theta$
between sample surface ($ab$-plane) and the magnetic field can be continuously changed, with currents flowing in the $ab$-plane perpendicular to magnetic field (inset in Fig. 2(a)). The magnetoresistance of SrMnBi$_2$ exhibits significant angular dependence (Fig. 2(a,b)). When $B$ is parallel to the $c$-axis ($\theta=0^{o}$), the MR is maximized and is linear above a characteristic field ($\sim1$ T). With increase in the tilt angle $\theta$, MR gradually decreases and becomes nearly negligible for $B$ in the $ab$-plane ($\theta=90^o$).

%%%%%%%%%%%%%%%%%%%%%%% Figure 2 %%%%%%%%%%%%%%%%%%%%%%%%%%%%%%%%%%
\begin{figure}[tbp]
\includegraphics[scale=0.8] {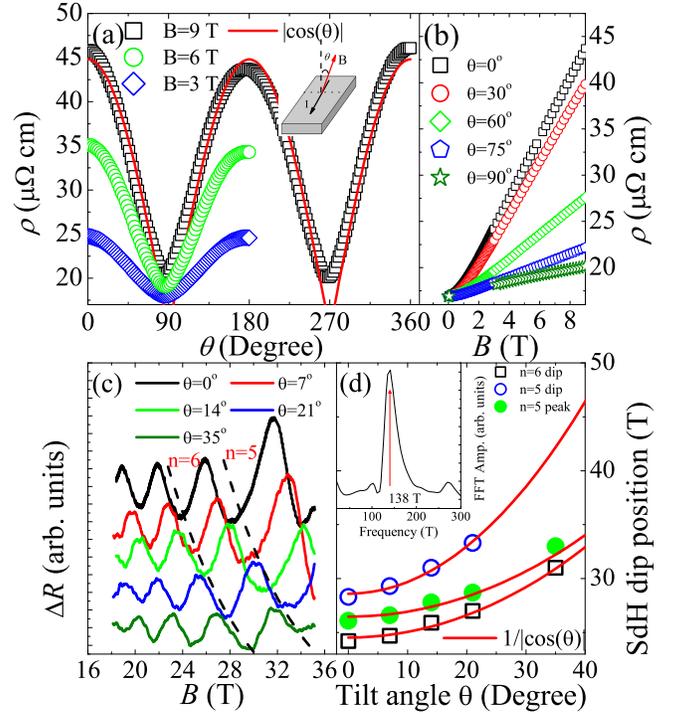}
\caption{(a) In-plane resistivity $\rho$ vs. the tilt angle $\protect\theta$ from $0^o$ to $
360^o$ at $B$ = 3, 6 and 9 T and $T$ = 2 K. The red line is the fitting curve using $|\cos(
\protect\theta)|$ (see text). Inset shows the configuration of the measurement. (b) In-plane Resistivity $\protect\rho$ vs. magnetic field $B$ with different tilt angles $\protect\theta$ at 2 K. (c) MR SdH oscillations $\Delta R=R_{xx}-<R>$ as a function of field $B$ below 35 T with tilt angles $\theta$ from $0^o$ to $35^o$ at 2 K. The dashed lines indicate the SdH dips at Landau filling factor $v=5$ and 6. The different curves are offset for clarification. (d) Position of dips with $v=5$ and 6, as well as the position of peaks with $v=5$, plotted against the tilt angle $\theta$. The data are consistent with the $1/|\cos(\theta)|$ dependence (red lines). The inset shows the Fourier transform of the SdH oscillation which gives a single frequency of $F=138(7)$ T.}
\end{figure}
%%%%%%%%%%%%%%%%%%%%%%%%%%%%%%%%%%%%%%%%%%%%%%%%%%%%%%%%%%%%%%%%%%%%

The response of the carriers to the applied magnetic field and the magnitude of MR is determined by the mobility in the plane perpendicular to the magnetic field.\cite{metal}
For nearly isotropic three-dimensional (3D) FS, there should be no significant angle-dependent MR (AMR). In (quasi-)2D systems, 2D states will only respond to perpendicular component of the magnetic field $B|\cos(\theta)|$ and consequently longitudinal AMR and AMR-oscillation were observed in some quasi-2D conductors (such as Sr$_2$RuO$_4$ and $\beta-$(BEDT-TTF)$_2$I$_3$) and the surface states of TIs.\cite{amro1,amro2,qt1,qt2} For example, the MR of the bulk state in topological insulator has only $\sim 10\%$ angular dependence while the angular dependence of MR in the surface state is about ten times larger.\cite{qt2} Significant AMR was also observed in some materials with highly anisotropic 3D FS such as Bi and Cu, but the period of AMR is determined by the shape of the Fermi surface and is very different from the one in 2D systems. In Bi, electrons exhibit a threefold valley degeneracy and in-plane mass anisotropy, so the AMR peaks each time when the magnetic field is oriented along the bisectrix axis and has a 60$^o$ period.\cite{bi} In Cu the AMR is more complex due to the complex FSs and peaks about every 25$^o$.\cite{cu}

The electronic structure calculations in SrMnBi$_2$ show that the states near the Fermi energy $E_F$ are dominated by the Bi states in the Bi square nets. Consequently the dominant FS should be quasi-2D. Angular dependent resistivity in $B=9$ T and $T=2$ K shows wide maximum when the field is parallel to the $c$-axis ($\theta=0^o, 180^o$), and sharper minimum around $\theta=90^o, 270^o$ (Fig. 2(a)). The whole curve of AMR in SrMnBi$_2$ follows the function of $|\cos(\theta)|$ very well with a 180$^o$ period (red line in Fig. 2(a)).  Moreover, the larger $\rho_{c}$ than $\rho_{ab}$ in Fig. 1(b) implies that the transfer integral and the coupling between layers along the $c$-axis is very small. All this implies that the FSs in SrMnBi$_2$ should be highly anisotropic and that the mobility of carriers along $k_z$ is much smaller than the value in $k_xk_y$ plane.

Angular dependent MR quantum oscillations are directly related to the cross section of FS. In Fig. 2(c), the in-plane $\Delta R=R-<R>$ measured using same configuration as shown in the inset of Fig. 2(a) exhibits clear Shubnikov-de Hass (SdH) oscillations with tilt angles $\theta$ from $0^o$ to $35^o$. In metals, SdH oscillations correspond to successive emptying of LLs by the magnetic field and the LLs index $n$ is related to the cross section of FS $S_F$ by $2\pi(n+\gamma)=S_F\frac{\hbar}{eB}$.\cite{osc1,qt1,qt2} For a 2D FS (a cylinder), the cross section has $S_F(\theta)=S_0/|\cos(\theta)|$ angular dependence and the LLs positions should be inversely proportional to $|cos(\theta)|$.\cite{qt1,osc1} The peak (dip) positions in SrMnBi$_2$ rapidly shift toward higher field direction with increase in $\theta$ (as indicated by the dashed lines in Fig. 2(c)). In Fig. 2(d), the dip positions corresponding to LLs $n=5, 6$ and the peak position with $n=5$ were plotted against the tilt angle $\theta$ and can be described very well by $1/|\cos(\theta)|$ (the red lines in Fig. 2(d)). Similar behavior was observed in the surface states of TIs~\cite{qt1,qt2} and some other layered structures.\cite{osc1,osc2} \textbf{The Fourier transform of the SdH oscillation (inset of Fig. 2(d)) revealed that the oscillation component shows a periodic behavior in $1/B$ with a single frequency $F=138(7)$ T. The small value of frequency is consistent with previous value in Ref.[7] and demonstrates that the dominant FSs are very small since the Onsager relation is $F=(\Phi_0/2\pi^2)A_k$ where $\Phi_0$ is the flux quantum and $A_k$ is the cross sectional area of FS.\cite{osc1} This clearly shows that the dominant two-dimensional FSs found in Fig. 2(b) are indeed the small FSs between $\Gamma$ and $M$ points, rather than the large FS at $\Gamma$ point in the Brillouin zone.} Above SdH oscillation combined with the angular MR clearly suggests that the dominant FSs of SrMnBi$_2$ are small quasi-2D cylinders along $k_z$ originating from Bi square nets. In addition, there are still conventional parabolic bands with three dimensional characteristic close to the Fermi level,\cite{srmnbi21} causing the small deviation from quasi-2D transport.

%%%%%%%%%%%%%%%%%%%%%%% Figure 3 %%%%%%%%%%%%%%%%%%%%%%%%%%%%%%%%%%
\begin{figure}[tbp]
\includegraphics[scale=0.65] {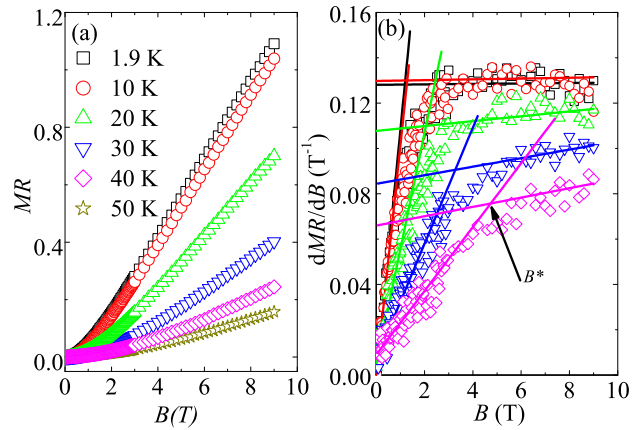}
\caption{(a) The magnetic field ($B$) dependence of the in-plane magnetoresistance (MR $=(
\protect\rho(B)-\protect\rho(0))/\protect\rho(0)$) at different
temperatures. (b) The field derivative of in-plane MR at different temperature
respectively. The lines in high field regions were fitting results using MR $=A_1B+O(B^2)$ and the lines in low field regions using MR $=A_2B^2$. }
\end{figure}
%%%%%%%%%%%%%%%%%%%%%%%%%%%%%%%%%%%%%%%%%%%%%%%%%%%%%%%%%%%%%%%%%%%%

%%%%%%%%%%%%%%%%%%%%%%% Figure 4 %%%%%%%%%%%%%%%%%%%%%%%%%%%%%%%%%%
\begin{figure}[tbp]
\includegraphics [scale=1]{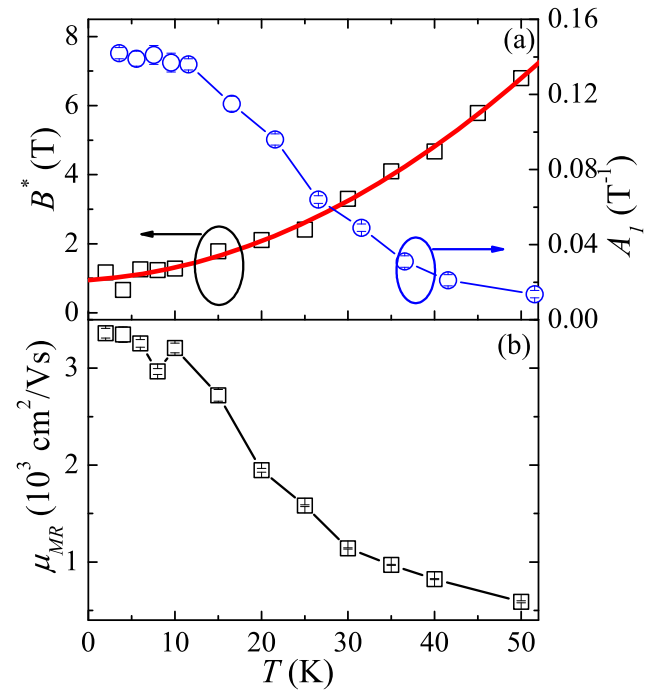}
\caption{(a) Temperature dependence of the critical field $B^*$ (black
squares) and the high field MR linear coefficient $A_1$ (blue circles) up to
50 K. The red solid line is the fitting results using $B^*=\frac{1}{2e\hbar
v_F^2}(E_F+k_BT)^2$. (b) Temperature dependence of the
effective MR mobility $\protect\mu_{MR}$ extracted from the weak-field MR.}
\end{figure}
%%%%%%%%%%%%%%%%%%%%%%%%%%%%%%%%%%%%%%%%%%%%%%%%%%%%%%%%%%%%%%%%%%%%

Now we turn to the linear nonsaturated in-plane magnetoresistance in SrMnBi$_2$ (Fig. 3(a)). The MR
is linear over a wide field range up to 50 K. This behavior extends to very low fields
where the MR naturally reduces to a weak-field semiclassical quadratic
dependence. The cross-over from the weak-field $B^2$ dependence to the
high-field linear dependence can be best seen by considering the field
derivative of the MR, $d$MR$/dB$ (Fig. 3(b)). In the low field range ($B<$1 T at 2 K), $d$MR$/dB$ is proportional to $B$ (as shown by lines in low-field regions),
indicating the semiclassical MR $\sim A_2B^2$. But above a
characteristic field $B^*$, $d$MR$/dB$ deviates from the
semiclassical behavior and saturates to a much reduced slope (as
shown by lines in the high-field region). This indicates that the MR for $B>B^*$ is dominated
by a linear field dependence plus a very small quadratic term (MR$
=A_1B+O(B^2)$).

The linear MR deviates from the semiclassical $B^2$ dependence of MR in the low field
region and a saturating MR in high fields.\cite{metal} The unusual nonsaturating linear magnetoresistance has been reported in gapless semiconductor Ag$
_{2-\delta}$(Te/Se) \cite{agte1,agte2} with the linear energy spectrum in the quantum limit.\cite
{quantummr,agte2} Recent first principle calculations confirmed that these
 materials have a gapless Dirac-type surface state.\cite{fangzhong}
Linear magnetoresistance is also observed in topological
insulators \cite{qt1,qt2} and BaFe$_2$As$_2$ \cite{qt3} with Dirac fermion states. Another possible origin of the large linear magnetoresistance is the
mobility fluctuations in a strongly inhomogeneous system.\cite{disorder} This does not apply in SrMnBi$_2$
since our sample is stoichiometric crystal without doping/disorder. Below we show that the nonsaturating linear magnetoresistance and the deviation from the semiclassical transport in SrMnBi$_2$ is due to the linear energy dispersion.

The application of a strong
perpendicular external magnetic field ($B$) would lead to a complete
quantization of the orbital motion of carriers with linear energy dispersion, resulting in quantized LLs $
E_n=sgn(n)v_F\sqrt{2e\hbar B|n|}$ where $n=0,\pm1,\pm2,\cdots$ is the LL
index and $v_F$ is the Fermi velocity.\cite
{LL1,LL2} Then the energy splitting between the lowest and $1^{st}$ LLs is
described by $\triangle_{LL}=\pm v_F\sqrt{2e\hbar B}$.\cite{LL1,LL2} In the
quantum limit at specific temperature and field, $\triangle_{LL}$ becomes
larger than both the Fermi energy $E_F$ and the thermal fluctuations $k_BT$
at a finite temperature. Consequently all carriers occupy the lowest
Landau level and eventually the quantum transport with linear magnetoresistance shows
up. The critical field $B^*$ above which the quantum limit is satisfied at
specific temperature $T$ is $B^*=\frac{1}{2e\hbar v_F^2}
(E_F+k_BT)^2$.\cite{qt3} The splitting of LLs in conventional
parabolic bands is $\triangle_{LL}=\frac{e\hbar B}{m^*}$. Hence the
evolution of $\triangle_{LL}$ with field for parabolic bands is much slower
than that for Dirac fermion states, and it is difficult to observe
quantum limit behavior in the moderate field range. The
temperature dependence of critical field $B^*$ in SrMnBi$_2$ clearly
deviates from the linear relationship and can be well fitted by $B^*=\frac{1}{
2e\hbar v_F^2}(E_F+k_BT)^2$, as shown in Fig. 4(a). The fitting gives the Fermi velocity $v_F\sim 5.13\times 10^5$ ms$^{-1}$ and $\Delta_1\sim 4.97$ meV. This confirms the existence of Dirac fermion states in SrMnBi$_2$.

In a multiband system with both Dirac and conventional
parabolic-band carriers (including electrons and holes) the
magnetoresistance in the semiclassical transport can be described as $MR=
\frac{\sigma_e\sigma_h(\mu_e+\mu_h)^2}{(\sigma_e+\sigma_h)^2}B^2$ where $
\sigma_e, \sigma_h, \mu_e, \mu_h$ are the effective electron and hole
conductivity and mobility in zero field respectively, when the Dirac
carriers are dominant in transport.\cite{qt3,qt4} Then the coefficient of
the low-field $B^2$ quadratic term, $A_2$, is related to the effective MR
mobility $\sqrt{A_2}=\frac{\sqrt{\sigma_e\sigma_h}}{\sigma_e+\sigma_h}
(\mu_e+\mu_h)=\mu_{MR}$, which is smaller than the average mobility of carriers $\mu_{ave}=\frac{\mu_e+\mu_h}{2}$ and gives an estimate of the lower bound. Fig. 4(b) shows the dependence of $\mu_{MR}$ on the
temperature. At 2 K, the value of $\mu_{MR}$ is about 3400 cm$^2$/Vs. The large effective MR mobility also implies that Dirac
fermions dominate the transport behavior. With increase in temperature, the value of $\mu_{MR}$ and the coefficient of high-field linear term $A_1$ (Fig. 4(a)) decrease sharply. This is due to thermal fluctuation smearing out the LL splitting.

In summary, we demonstrate quantum transport of 2D Dirac fermion states in bulk SrMnBi$_2$ single crystals. The bands with linear energy dispersion lead to the large nonsaturated linear magnetoresistance since all Dirac fermions occupy the lowest Landau level in the quantum limit. The transverse magnetoresistance exhibits a crossover at a critical field $B^*$ from semiclassical weak-field $B^2$ dependence to the high-field linear-field dependence. With increase in temperature, the critical field $B^*$ increases and the temperature dependence of $B^*$ satisfies quadratic behavior which is attributed to the Landau level splitting of the linear energy dispersion. The effective magnetoresistant mobility $\mu_{MR}\sim 3400$ cm$^2$/Vs comparable to values observed in graphene is observed. The angle-dependence of magnetoresistance shows a $|\cos(\theta)|$ dependence while the LL positions in SdH oscillations are inversely proportional to $|\cos(\theta)|$, indicating the dominant quasi-2D Fermi surfaces. Our results show that the crystals with Bi square nets can host phenomena commonly observed so far in 2D structures and materials like graphene.

%\begin{acknowledgments}
We than John Warren for help with SEM measurements. Work at Brookhaven is supported by the U.S. DOE under contract No. DE-AC02-98CH10886. Work at the National High Magnetic
Field Laboratory is supported by the DOE NNSA DEFG52-10NA29659 (S. W. T and D. G.), by the NSF Cooperative Agreement No. DMR-0654118 and by the state of Florida.
%\end{acknowledgments}

% Create the reference section using BibTeX:
%\bibliography{basename of .bib file}

\end{document}